\documentclass{aipproc}
\usepackage{epsf}  
\layoutstyle{6x9}

\newcommand{\beq}{\begin{eqnarray}}
\newcommand{\eeq}{\end{eqnarray}}
\newcommand{\s}{\bm{\sigma}}

\newcommand{\p}{\bm{p}}
\newcommand{\hp}{\bm{\widehat{\p}}}
\newcommand{\x}{\bm{x}}
\newcommand{\0}{\bm{0}}
\newcommand{\bv}{\bm{\varphi}}

\newcommand{\hbv}{\bm{\widehat\varphi}}

\newcommand{\bl}{\bm{\lambda}}
\newcommand{\br}{\bm{\rho}}

\begin{document}

\title{ Lagrangians for invariant sub-spaces of the squared 
Pauli-Lubanski vector }

\author{G. Cabral$^1$, and M. Kirchbach$^2$}
{address={$^1$Facultad de Fis\'{\i}ca, Univ.\ Aut.\ de Zacatecas,
A.P. C-600, Zacatecas, ZAC-98062, M\'exico\\
$^2$Instituto de F\'{\i}sica, 
         Univ.\ Aut.\ de San Luis Potos\'{\i},
     San Luis Potos\'{\i}, S.L.P. 78240, M\'exico}}

\begin{abstract}
We present an alternative formalism to the Rarita-Schwinger
framework for the description of ``has-been'' higher-spins
at rest that avoids the Velo-Zwanziger problem.
\end{abstract}

\maketitle

\def\beq{\begin{eqnarray}}
\def\eeq{\end{eqnarray}}

\def\A{{\mathcal A}^\mu}
\def\W{{\mathcal W}_\mu}

\def\beq{\begin{eqnarray}}
\def\eeq{\end{eqnarray}}

\def\s{\mbox{\boldmath$\displaystyle\mathbf{\sigma}$}}
\def\p{\mbox{\boldmath$\displaystyle\mathbf{p}$}}
\def\hp{\mbox{\boldmath$\displaystyle\mathbf{\widehat{\p}}$}}
\def\x{\mbox{\boldmath$\displaystyle\mathbf{x}$}}
\def\0{\mbox{\boldmath$\displaystyle\mathbf{0}$}}
\def\bv{\mbox{\boldmath$\displaystyle\mathbf{\varphi}$}}
\def\hbv{\mbox{\boldmath$\displaystyle\mathbf{\widehat\varphi}$}}

\def\bl{\mbox{\boldmath$\displaystyle\mathbf{\lambda}$}}
\def\bl{\mbox{\boldmath$\displaystyle\mathbf{\lambda}$}}
\def\br{\mbox{\boldmath$\displaystyle\mathbf{\rho}$}}
\def\1{\openone}
\def\bfhh{\mbox{\boldmath$\displaystyle\mathbf{(1/2,0)\oplus(0,1/2)}\,\,$}}

\def\mn{\mbox{\boldmath$\displaystyle\mathbf{\nu}$}}
\def\amn{\mbox{\boldmath$\displaystyle\mathbf{\overline{\nu}}$}}

\def\mne{\mbox{\boldmath$\displaystyle\mathbf{\nu_e}$}}
\def\amne{\mbox{\boldmath$\displaystyle\mathbf{\overline{\nu}_e}$}}
\def\rlh{\mbox{\boldmath$\displaystyle\mathbf{\rightleftharpoons}$}}

\def\wm{\mbox{\boldmath$\displaystyle\mathbf{W^-}$}}
\def\hh{\mbox{\boldmath$\displaystyle\mathbf{(1/2,1/2)}$}}
\def\h00h{\mbox{\boldmath$\displaystyle\mathbf{(1/2,0)\oplus(0,1/2)}$}}
\def\znbb{\mbox{\boldmath$\displaystyle\mathbf{0\nu \beta\beta}$}}

\noindent
\underline{\it Introduction.}
One of the long standing problems in particle physics is the covariant
description of higher spin states. The standard formalism is based upon 
totally symmetric Lorentz invariant tensors of rank-K with Dirac
spinor components, $\psi_{\mu_1...\mu_K}$, which satisfy the Dirac equation
for each space-time index. In addition, one requires 
$\partial^{\mu_1}\psi_{\mu_1...\mu_K}=0$, and 
$\gamma^{\, \mu_1}\psi_{\mu_1...\mu_K}=0$. The solution obtained this way
(so called Rarita-Schwinger framework) describes 
a ``has--been'' spin-$(K+{1\over 2})$ particle at rest that is a parity
singlet, i.e. without a companion of opposite parity, 
and a particle of indefinite spin in flight. 
Problems occur when $\psi_{\mu_1...\mu_K}$
constrained this way
are placed within an electromagnetic field. In this case, the energy
of the ``has-been'' spin-$(K+{1\over 2})$ state becomes imaginary,
and it propagates acausally (Velo-Zwanziger problem [1]).
Here we first emphasize that the
$\gamma^{\, \mu_1}\psi_{\mu_1...\mu_K} =0$ constraint
is a short-hand of the definition of the 
above parity singlet as one of the
invariant subspaces of the squared Pauli-Lubanski vector, $W^2$. 
We consider the simplest case of $K=1$ and 
construct the covariant projector onto that very state 
as $-{1\over 3}({1\over m^2}W^2 +{3\over 4})$.
We suggest to work in the sixteen dimensional vector
space, $\Psi$, of the direct product of the four-vector, $A_\mu$,
with the Dirac spinor, $\psi$,
i.e.  $\Psi =A \otimes \psi  $, rather than keeping 
space-time and spinor indices separated (as in $\psi_\mu$)
and to  consider 
$\left(- {1\over 3}\left({1\over m^2}W^2 +{3\over 4}\right)-1 \right)\Psi =0$ 
as the principal wave equation without invoking any further supplementary 
conditions. In gauging the equation minimally, calculating and nullifying
its determinant, we obtained the energy-momentum dispersion relation.
The latter turned out to be well behaved and free of pathologies,
thus avoiding the classical Velo-Zwanziger problem.

\noindent
\underline{\it Projectors onto $W^2$ invariant sub-spaces.}
The Pauli--Lubanski vector is defined as 
\begin{equation}
W_\mu  =- {1\over 2} 
\epsilon_{\mu\nu\rho\tau} S^{\nu\rho } P^\tau\, ,
\label{PauLu}
\end{equation}
where $\epsilon_{0123}=1$, and $P^\tau $ are the generators of translations.
Its squared is calculated to be
\begin{equation}
W^2= 
- {1\over 2} S\cdot S\, P^2
+ G\cdot G\, , \quad G^\nu =S^{\sigma \nu} P_\sigma\, .
\label{w2}
\end{equation}
In $(s,0)/(0,s)$, $G^2=-W^2$ and $W^2$ reduces to $-m^2\mbox{\bf S}^2$.
{}For fields of the type $(s_1,s_2)$ with $s_1$ and $s_2$ non-vanishing, 
$G^2\not=W^2$ and spin is no longer a good quantum label for 
such particles in flight. To build $W_\mu$ in $A_\nu\otimes \psi$ we 
recall generators of the Dirac representation 
\beq
S_{\nu\rho }^{\left({1\over 2},0\right)\oplus \left(0,{1\over 2}\right) }  =
{1\over 2}\sigma_{\nu\rho }\, ,&\quad&
\sigma_{\nu\rho}={i\over 2}\lbrack \gamma_\nu, \gamma_\rho \rbrack\, ,  
\label{Dirsmn}
\eeq
where $\gamma_\mu $ are the Dirac matrices.
The Lorentz generators in the four-vector 
$\left({1\over 2},{1\over 2}\right)$
space are obtained from those of the right-handed  
$\left( {1\over 2},0\right)$, and left-handed
$\left( 0, {1\over 2}\right)$ spinors [2-4]
in noticing that $\left({1\over 2},{1\over 2}\right)$ is the direct product
of $\left( {1\over 2},0\right)$ and $\left( 0, {1\over 2}\right)$. 
One finds [4]
\beq
S^{\left( {1\over 2}{1\over 2}\right) }_{0 l} = 
1_2 \otimes (i\sigma_l) +(- i\sigma_l)\otimes 1_2\, ,
\quad
S^{\left( {1\over 2}{1\over 2}\right) }_{ij}=
i\epsilon_{ijk}(1_2\otimes \sigma_k +\sigma_k\otimes 1_2)\, , 
\label{spinor_gen}
\eeq
where $\sigma_k$ are the Pauli matrices.
Now the generators in $\psi_\mu /\Psi $ are cast into the form [3,4]:
\beq
S_{\nu\rho} =S^{\left( {1\over 2}{1\over 2}\right) }_{\nu \rho}\otimes 1_4 +
1_4\otimes {1\over 2} \sigma_{\nu \rho}\, .
\label{psi_mu_gen}
\eeq
In the finite dimensional representation space,
$\left( {1\over 2},{1\over 2} \right)$,
where one replaces the Poincar\'e generators of translations, $P_\mu$,
with their eigenvalues, $p_\mu$, i.e. by numbers,
$\Big[ W^{\left( {1\over 2}{1\over 2}\right) }\Big]\, ^2$ is no longer
proportional to the unit matrix and does not
covariantly commute with the operator of the squared spin, {\bf S}$^2$.
To account for this peculiarity ~[4], notation is
changed to $\widetilde{W}^{\left({1\over 2}{1\over 2}\right)}_\mu $.
As an immediate application, 
the corresponding $\widetilde{W}^2$ in the product space, be it in
$\psi_\mu$, or, $\Psi$, splits
these spaces into covariant sectors corresponding to 
different $\widetilde{W}^2$ eigenvalues according to 
\beq
\left( 
\widetilde{W}^{\left( {1\over 2}{1\over 2}\right) }\otimes 1_4 +
1_4\otimes W^{ ({1\over 2},0)\oplus (0,{1\over 2})}
\right)^2  \Psi^{\tau_l} & = &
-\tau_l\,  m^2\,\Psi^{\tau_l}\,, \,\,\,
\tau_l=(l\pm {1\over 2})(l\pm {1\over 2}+1)\, .
\label{inv_subsp}
\eeq
Because of $[\widetilde{W}^2,${\bf S}$^2]\not= 0$,
the $\tau_l$  sectors (with $l=1,0$) 
describe boosted ``{\it has--been\/}'' spin-$ (l\pm {1\over 2})$ 
states in the rest frame and states of undetermined spin in flight. 
On mass shell,  $p^2= m^2$,  
Eq.~(\ref{inv_subsp}) can be cast into the form
(compare Ref.~[3]) 
\begin{eqnarray}
P^{{{15}\over 4}}   (\p) \Psi^{{{15}\over 4}}   (\p) = 
\Psi  ^{{{15}\over 4}} (\p) \, , &\, &
P^{{{15}\over 4}}   (\p)  = -{1\over {3 }}\left[ 
{1\over m^2} \widetilde{W}^2+{3\over 4}  
\left( 1_4\otimes 1_4\right) \right]\, .
\label{projs1}
\eeq
It is easily verified that the operators $P^{{{15}\over 4}}(\p ) $, and
$P^{{3\over 4}}(\p )=1_{16}-P^{{{15}\over 4}}(\p ) $ are 
covariant projectors onto the $\widetilde{W}^2$ invariant subspaces with 
$\tau_1 ={{15}\over 4} $, which is a parity singlet,
and the two states of opposite parities
with $\tau_1=\tau_0={3\over 4}$, respectively, i.e.
\beq
{\Big[} P^{{{15}\over 4}} (\p){\Big]} ^2 = 
P^ {{{15}\over 4}} (\p) \, , &\quad& 
{\Big[} P^ {{{3}\over 4}}  (\p){\Big]}  ^2 = P^ {{{3}\over 4}}  (\p) \, , 
\quad P^{{{15}\over 4}} (\p)\, P^{{{3}\over 4}} (\p) =0\, .
\label{FNProj}
\eeq
In Ref.~[5] we showed that the second auxiliary condition
in the Rarita-Schwinger framework, $\gamma^{\, \mu}\psi_\mu=0$, 
is a short-hand from  Eq.~(\ref{projs1}) and is determined as
\beq
{1\over {m\alpha_{l} }}\gamma^{\, \epsilon} 
\left( {\widetilde W}^{\left({1\over 2},{1\over 2}\right )}_
{\epsilon \eta }\cdot \gamma\gamma_5 \right)
 \left(\psi^{\tau_l} \right) ^\eta (\p) &=&  
\gamma\cdot \, \psi^{\tau_l} (\p) \, .
\label{Step2}
\eeq
Here, $\alpha_{l}  $ is the eigenvalue of 
${\Big[} \widetilde{W}^
{ \left( {1\over 2} {1\over 2}\right)}\otimes 1_4{\Big]}\cdot 
{\Big[} 1_4\otimes W^{({1\over 2},0)\oplus (0,{1\over 2})}{\Big]}$ 
with respect to $\psi_\mu ^{ \tau_1 }$.
Upon using Eq.~(\ref{w2}), and $p^2=m^2$, 
Eq.~(\ref{FNProj}) can be cast into the form
\beq
P^{{{15}\over 4}}   (\p)  &=& -{1\over 3 }\left[ 
- {1\over 2}S\cdot S +  {1\over m^2} G\cdot G
+ {3\over 4}  \left( 1_4\otimes 1_4\right)  \right]\, .
\label{FNProj_const}
\eeq

\noindent
\underline{\it Propagators and Lagrangians for $\widetilde{W}^2$ invariant
sub-spaces.}
In having favored the 16 dimensional vector column space $\Psi $ 
over $\psi_\mu$, we gained the
advantage that the Dirac equation has been automatically accounted
for by means of the definition of the
Lorentz generators within $\Psi $,
i.e. through the second term on the rhs  in Eq.~(\ref{psi_mu_gen}).
The propagator associated with  
Eqs.~(\ref{projs1}), (\ref{FNProj_const}) is now deduced as the following 
$16\times 16$  matrix:
\begin{eqnarray}
{\mathcal S}^{{{15}\over 4 }}(\p) &= &
 { {\widetilde{P}^{{ {15}\over 4}} (\p) }\over 
{p^2-m^2}}\, =  {{ 2S\cdot S- {4\over m^2}G\cdot G
- 3\,\,  \left(1_4\otimes 1_4\right)  }\over 
{12 (p^2-m^2)} 
}\, .\nonumber\\ 
\label{FNProp}
\end{eqnarray} 
It directly verifies that this propagator corresponds to the 
following Lagrangian
\begin{eqnarray}
{\mathcal L}^{{{15}\over 4}} &= & \bar{\Psi}^{{{15}\over 4}}\, 
 {\Big[} 
2m^2 S^{\mu\nu} S_{\mu\nu}  -
4 S_{\mu\nu} p^\mu S^{\rho\nu }p_\rho {\Big]}
 \Psi^{{{15}\over 4}} -  15  m^2 \, \left(1_4\otimes 1_4\right) 
\bar{\Psi}^{ { {15}\over 4} } \Psi^{{{15}\over 4}} \, .
\label{lagr}
\end{eqnarray} 

\noindent
\underline{\it Energy-momentum dispersion relations
in the presence of  electromagnetic fields. }
We here studied the energy--momentum dispersion relation 
of Eq.~(\ref{projs1}) in the presence of
a simple magnetic field 
oriented along the $z$ axis, here denoted by $B_z$. We also took for
the sake of simplicity of the calculation the $z$ axis along $\p $. 
With the help of the symbolic code Mathematica we
then calculated  the appropriate determinant and, in nullifying it,
found the energy-momentum dispersion relation to be
\beq
E^2=(p_z -e  B_z )^2 + m^2\, ,
\label{gauge_disp}
\eeq
and therefore free of the Velo-Zwanziger problem 
of complex energy in the background of a magnetic field.
Therefore, the associated  interacting propagators can now be obtained 
in the standard way by replacing $/\!\!\!\!\!\!p$ through
$/\!\!\!\!\!\pi :=(p^\mu-eA^\mu)\gamma_\mu $. 
In having done so, we have produced a 
{\it  pathology-free  propagating $\tau_1 ={{15}\over 4} $ sector \/}
in the presence of an electromagnetic field.
Admittedly, we did not produce pure-spin propagators.
Nonetheless, we created a formalism that at least 
allows for the covariant description of the propagating 
``has-been'' spin-${3\over 2}^-$ piece of the polar vector spinor. 
\section{References}

\noindent

\noindent
$1.$ G.\ Velo and D.\ Zwanziger, 
 Phys.\ Rev.\ {\bf 186}, 1337-1342 (1969). 

\noindent
$2.$ L.\ H.\ Ryder, {\it Quantum field theory\/} 
(Cambridge Univ.\ Press, Cambridge,1987).

\noindent
$3.$ M.\ Kirchbach, 
Mod.\ Phys.\ Lett.\ {\bf A12}, 3177-3188 (1997). 

\noindent
$4.$ M.\ Kirchbach and D.\ V.\ Ahluwalia,
Phys.\ Lett.\ {\bf B529}, 124-131 (2002).

\noindent
$5.$ M.\ Kirchbach and D.\ V.\ Ahluwalia,
{\tt hep-ph/0210084}.

\end{document}